\documentstyle[11pt, epsf]{article}

\newcommand{\be}{\begin{equation}}
\newcommand{\ee}{\end{equation}}
\newcommand{\bea}{\begin{eqnarray}}
\newcommand{\eea}{\end{eqnarray}}
\newcommand{\vp}{\varphi}
\newcommand{\del}{\partial}

\renewcommand{\(}{\left(}
\renewcommand{\)}{\right)}
\renewcommand{\[}{\left[}
\renewcommand{\]}{\right]}
\newcommand{\eqn}[1]{(eq.~(\ref{#1}))}
\newcommand{\eq}[1]{eq.~(\ref{#1})}

\newcommand{\Tr}{{\rm Tr}\,}
\newcommand{\cvec}[2]{{#1 \choose #2}}
\begin{document}

\begin{titlepage}
\begin{flushright}
HD--THEP--95--42\\
hep-ph/9509352
\end{flushright}
\vspace{1.5cm}
\begin{center}
\mbox{\bf\LARGE Bound States in the Hot Electroweak Phase}\\
\vspace{1cm}
{\bf
Hans-G\"unther Dosch\\
\vspace{.2cm}
Jochen Kripfganz\footnote{supported by Deutsche
Forschungsgemeinschaft}\\
\vspace{.2cm}
Andreas Laser\footnote{supported by Landesgraduiertenf\"orderung
Baden-W\"urttemberg\\[1ex]
e-mail addresses:\\[0.5ex]
\begin{tabular}{r}
  H.G.Dosch@thphys.uni-heidelberg.de\\
J.Kripfganz@thphys.uni-heidelberg.de\\
    A.Laser@thphys.uni-heidelberg.de\\
M.G.Schmidt@thphys.uni-heidelberg.de
\end{tabular}}\\
\vspace{.2cm}
Michael G.~Schmidt\\
}
\vspace{1cm}
Institut  f\"ur Theoretische Physik\\
Universit\"at Heidelberg\\
Philosophenweg 16\\
D-69120 Heidelberg, FRG\\
\vspace{1.5cm}
{\bf Abstract}\\
\end{center}
The high temperature phase of the electroweak standard theory is
described by a strongly coupled $SU(2)$-Higgs-model in three dimensions.
As in the Abbott-Farhi-model Higgs and $W$-boson are low lying bound
states. Using a method by Simonov based on the
Feynman-Schwinger representation of correlators we
calculate the masses of these states. Our results are compared with
lattice masses.
\end{titlepage}

The electroweak standard theory successfully predicted massive vector
bosons making use of the Higgs mechanism. Much of its strength is based
on the possibility to do rigorous perturbative calculations.
Later on interest also turned towards the behavior of this theory at
high temperature that has been realized in the early universe.
The proposal that there exists a hot phase where the Higgs field
obtains a positive thermal mass and the
\mbox{$SU(2)_L\!\otimes\! U(1)$}-symmetry
is restored  \cite{KirzhnitsLi} is commonly accepted. Investigations
of the electroweak phase transition have to treat the  hot and the cold
phase at the transition temperature in an appropriate way.

Perturbation theory in the broken phase has been studied extensively
\cite{AndersonHa}-\cite{Wir2}. For
\mbox{$\lambda \vp^4$-couplings} corresponding to zero temperature Higgs masses
below 70 GeV it works very well. A perturbative treatment of the hot
phase would predict a vanishing Higgs vacuum expectation value
and a massless gauge boson.
However perturbation theory alone is not a good guide in this phase due to
IR-problems.

The electroweak standard theory  at the transition temperature
allows the high temperature
expansion. To good accuracy it can be described by an effective 3-dimensional
$SU(2)$-Higgs-model whose parameters can be calculated from the fundamental
4-dimen\-sion\-al theory \cite{KajantieEA}.
All IR-problems appearing in the unbroken phase are
represented by this effective theory. It might be compared to
3-dimensional QCD but with a fundamental
scalar field $~\Phi~$ instead of quarks
and a $SU(2)$-gauge group.
$\Phi~$ corresponds to the perturbative Higgs particle. We avoid this name to
preclude confusions. Its mass $m_3$ will be called Lagrangian mass.

As in QCD the most secure way to calculate the properties of such a system
is a Monte-Carlo simulation
on the lattice. This is done intensely and first results are
appearing \cite{BunkEA}-\cite{IlgenfEA}.
Of course being aware of the long way to reliable hadron lattice
results one has to question if the present lattice sizes are appropriate
for measuring certain quantities.
An interesting alternative way to such 3-dimensional systems at high
temperature is the exact renormalization group approach \cite{BeWe}.
It requires, however, simplifying assumptions for practical
use, and these are not easily controlled.
Recently it was speculated from the investigation of 1-loop gap-equations
in the two settings with small and large Higgs vacuum expectation value,
that the hot phase is just another Higgs phase with different
parameters \cite{BuPh}. But this does not agree well with the
masses $m_W$ and $m_H$ observed on the 3-dimensional
lattice \cite{BunkEA,FodorEA,IlgenfEA}.
The subject of this letter is to propose a model which explains these
correlation masses; we compare with the lattice results at the end.\\
\rule{0cm}{0cm}

With rising 3-dimensional gauge coupling in the infrared \cite{ReuWe}
it is reasonable to propose  confinement like in QCD.
As in the Abbott-Farhi-model \cite{AbFa} physical states are bound
states and singlets with respect to the gauge group. They may nevertheless
be in a nontrivial isospin representation. Let us start with the
$SU(2)$-Higgs-model Lagrangian
\be\label{L}
{\cal L} \;=\; -\,\frac{1}{2}\, \Tr F_{kl}F^{kl}
    + (D_k \phi)^\dagger  (D^k \phi)
    - V(\phi^\dagger \phi) \quad.
\ee
$D_k = \del_k - ig A_k~~$ is the covariant derivative with the
gauge field
\mbox{~~$A_k = A_k^i\,\frac{\tau^i}{2}$\rule[-0.2cm]{0.0cm}{0.3cm}~~}
($\tau^i$ are the Pauli matrices),
$~~F_{kl} = [D_k, D_l]~~$ is the field strength tensor and
\mbox{$~~\phi= \cvec{\vp_1}{\vp_2}$ \rule[-0.2cm]{0.0cm}{0.3cm}~~}
is the scalar isospin doublet.
${\cal L} $ is invariant under the gauge transformation $~U(x)\in SU(2)$
\be
D_k \;\rightarrow\; U D_k U^\dagger  \qquad\qquad
\phi \; \rightarrow\; U \phi \quad.
\ee

The scalar doublet can unambiguously be written as
\mbox{~$\phi = \Phi \cvec{0}{1}$\rule[-0.2cm]{0.0cm}{0.3cm}},
where the $2\!\times \! 2$-matrix $~\Phi~$ is a linear combination of
$~1_{2\times 2}~$ and $~\tau^i$, and
\mbox{~~$\frac{1}{2}\Tr\, \Phi^\dagger\Phi =
              \phi^\dagger\phi$\rule[-0.2cm]{0.0cm}{0.3cm}~}.
Using this notation uncovers the $SU(2)\!\otimes\! SU(2)_I$ -invariance of
the Lagrangian \eqn{L}, namely $~~\Phi \rightarrow U \Phi V~$,
of which the left
$SU(2)$ is gauged while the isospin $SU(2)_I$ is global.

The confining theory has (among others) bound states corresponding to
the local interpolating fields
\be\label{intfields}
\Tr\,\Phi^\dagger\Phi \qquad {\rm and} \qquad
\Tr\,\Phi^\dagger D_k \Phi\, \tau^i \quad.
\ee
The first one is a scalar isospin singlet and is identified with the
``Higgs particle''.  The second one is an vector isospin triplet,
the ``$W$-boson''.
The scalar triplet operator $~~\Tr\,\Phi^\dagger \Phi \,\tau^i~~$
vanishes.
The operator $~~\Tr\,\Phi^\dagger D_k \Phi~~$
does not correspond to a vector singlet bound state but is
identical to $~~\del_k \Tr\,\Phi^\dagger\Phi~~$ and
generates the scalar singlet at non-zero momentum.
There are no bound states with these quantum numbers.

The nonlocal versions of the operators in \eq{intfields} are proportional to
\be\label{nlop}
\Tr\, \Phi(x)^\dagger T(x,\bar{x}) \Phi(\bar{x}) \qquad {\rm and} \qquad
\Tr\, \Phi(x)^\dagger T(x,\bar{x}) \Phi(\bar{x})\, \tau^i  \quad.
\ee
The space index of the vector operator is fixed by the direction of
the link operator $T$ (also known as gauge field transporter).
The latter is defined as (${\cal P}$ denotes the path ordering)
\be
T(x,\bar{x}) = {\cal P} \exp\(i g \int^{\bar{x}}_x A_k dz_k\) \quad.
\ee
These are the direct continuum counterparts of the lattice observables
interpreted as Higgs and $W$-boson.

It has been shown in the framework of sum rules, that this model reproduces
the phenomenology of the standard theory in the Higgs phase without any
use of naive spontaneous symmetry breaking but with an appropriate pattern
of $~~\Tr\,\Phi^\dagger\Phi~~$ vacuum structure \cite{SchDoKr}-\cite{BuSch}.
(There is no fundamental difference between the 3 and 4-dimensional case.)
Of course already at that time there was the other option (the
genuine Abbott-Farhi-model) that gluon condensates like in QCD lead
to a strongly bound phase \cite{Narison}.
Indeed this is the picture we propose for the
hot phase of electroweak theory.

The emerging physical picture would thus be very similar
to QCD in 3 space time dimensions,
but with spinless constituents, the fundamental scalar fields.
It is tempting to try to make connection to the confining
quark model of QCD which in a very
transparent way can explain many features of the bound state spectra.
Since the fundamental scalar particles are light
compared to the string tension
$\sigma$ we cannot use the simple non-relativistic quark model.\\
\rule{0cm}{0cm}

In the case of QCD Simonov \cite{Simonov} has used the Feynman-Schwinger
representation of hadronic correlators in order to treat light quarks kept
together by a linear confining potential. The method is easily adopted to the
case treated here, i.e.\ a theory in 3 space-time dimensions with scalar
``quarks''. In this case we even avoid the problems of the large
spin-interactions which lead presumably to zero modes and
chiral symmetry breaking.
Another difference is that the $W$-boson is an isospin-triplet, a state which
has no analogy in QCD.
Simonov's method is nevertheless applicable since the isospin structure
is not changed by the propagation.
The propagator of the matrix field
$\Phi^\dagger(x)_{a\alpha}$ to $\Phi(y)_{b\beta}$
is given in the Feynman-Schwinger representation by
($\alpha, \beta$ are gauge and $a, b$ are isospin indices)
\be\label{prop}
G(x,y)_{a \alpha b \beta} \;=\;
\int_0^\infty ds \exp\(-m_3^2 s\) \int{\cal D}z
\exp\(-\frac{1}{4} \int_0^s \dot{z}_k^2(\tau) d \tau \)
T(x,y)_{\alpha\beta} \,\delta_{ab} \quad.
\ee
The path integral $\int{\cal D}z$ runs over all curves $z(\tau)$
with tangent $\dot{z}(\tau)$ connecting $x$ and $y$.

Using a similar expression for the anti-scalar propagator from $\bar{x}$
to $\bar{y}$ one can put things together to get
the  Green function for a scalar--anti-scalar
state at position $(x,\bar{x})$
propagating to $(y,\bar{y})$ in the quenched approximation
\begin{eqnarray} \label{FSR}
G(x,\bar{x},y,\bar{y}) &=&  \int_0^\infty ds \int_0^\infty d\bar{s}
             \int \int {\cal D}z {\cal D}\bar{z}
           \exp\( -(m_3^2(s + \bar{s}) \) \;\;\times\\
&\times&\; \exp\( -\frac{1}{4}\int_0^s \dot{z}_k^2(\tau) d\tau
             -\frac{1}{4}\int_0^{\bar{s}} \dot{\bar{z}}_k^2(\bar{\tau})
         d\bar{\tau}\)
          {\cal P} \exp\(ig \oint_{x,\bar{x},y,\bar{y}} A_k dz_k\)
                 \nonumber \quad.
\end{eqnarray}
The last factor of \eq{FSR} is just the Wegner Wilson integral
over the closed loop formed by the gauge field transporters $T$
of the propagators \eqn{prop} and the bound state operators \eqn{nlop}.
In a first approach it is approximated by the area law
\be\label{arealaw}
{\cal P} \exp\( ig \oint_{x,\bar{x},y,\bar{y}} A_k d z_k\)
\;\propto\; \exp\(- \sigma F\) \quad,
\ee
where $F$ is the area and $\sigma$ the string tension.

We are evaluating the Green function \eq{FSR} following Simonov \cite{Simonov}.
Making some plausible assumptions on the (nearly) minimal surfaces
the surface $F$ is parameterized by $z_k(\tau)$ and
$\bar{z}_k(\tau)$.
In a next step the ``center of mass trajectory'' $R_k$  and the relative
coordinate $u_k$ are introduced
\begin{eqnarray}\label{Sub}
R_k \;=\; \frac{s \bar{s}}{s+\bar{s}}
          \( \frac{1}{s} z_k + \frac{1}{\bar{s}} \bar{z}_k\)
\qquad\quad
u_k \;=\; z_k - \bar{z}_k \quad.
\end{eqnarray}
The path integrals over  $z_k$ and $\bar{z}_k$ transform into
path integrals over  $R_k$ and $u_k$.

Making the assumption that the center of mass motion is dominated
by the classical path the path integral over $R_k$ can be replaced
by the parameterization $~R_k(\gamma)$~. Without loss of generality
we choose $~~R_1 = R_2 = 0~$, $~~R_3 = \gamma \Theta = \vartheta~~$
with $~~0 \le \gamma \le 1~$, where $\Theta$ is the distance between
the centers of
mass of the bounded scalar--anti-scalar system at position
$(x,\bar{x})$ and at position $(y,\bar{y})$.
This assumption is the most stringent one and should become better for
heavier states (as compared to $\sqrt{\sigma}$).

One then can rewrite the Green function \eqn{FSR} as
(for detail see ref.~\cite{Simonov})
\begin{equation}
G(x,\bar{x},y,\bar{y}) \;\propto\;
\int_0^\infty d s\int_0^\infty d\bar{s} \int {\cal D}u \exp(-B)
\end{equation}
with the three dimensional action
\begin{equation}\label{act}
B \;=\; \int_0^\Theta d\vartheta \[\; m_3^2 \( \frac{s+\bar{s}}{\Theta} \)
         + \frac{s+\bar{s}}{4 s \bar{s}} \Theta + \frac{\Theta}{4(s + \bar{s})}
           \( \frac{\partial \vec{u}}{\partial \vartheta} \)^2
         + \sigma \sqrt{u_1^2+u_2^2} \;\] \quad.
\end{equation}
It is convenient to introduce the new variables
\begin{equation}\label{mus}
\mu_1 = \frac{\Theta}{2 s} \qquad \mu_2 = \frac {\Theta}{2 \bar{s}} \qquad
\tilde{\mu} =  \frac{\mu_1 \mu_2}{\mu_1 + \mu_2} \quad.
\end{equation}
The $u_1$ and $u_2$ path integral can be substituted by the solutions of the
two dimensional Schr\"odinger equation
\begin{equation}\label{schr}
H \psi(u_1,u_2) = \epsilon \; \psi(u_1,u_2) \qquad \mbox{with} \qquad
H \;=\; - \,\frac{1}{2\tilde{\mu}} \( \frac{\partial^2}{\partial u_1^2}
          + \frac{\partial^2}{\partial u_2^2} \)
          + \sigma \sqrt{u_1^2+u_2^2} \quad.
\end{equation}
This is most easily seen if one interprets $B$ \eqn{act} as the
the Euclidian action of a point particle in two space and
one time dimension.

The last path integral over $u_3$ is now trivial and gives a negligible
contribution.

The remaining $s$ and $\bar{s}$ integrals are finally evaluated by the
method of steepest descend.
The correlation mass $M$ of the bound states, defined by
$~~G \propto \exp(-M\Theta)~$,
is hence found by minimizing
the integrand of $B$ evaluated at the solutions of the Schr\"odinger
equation with respect to $\mu_1$ and $\mu_2$. Since the Lagrangian
masses of the both fundamental scalars are the same ($m_3$) one
gets only one condition
\begin{equation}\label{steep}
\frac{\partial M(\mu)}{\partial \mu} = 0 \qquad \qquad
M(\mu) = \frac{m_3^2}{\mu} + \mu + \epsilon(\mu)
\end{equation}
with $~~\mu_1 = \mu_2 = \mu = 2\tilde{\mu}~$.
$\epsilon(\mu)$ is the eigenvalue introduced in \eq{schr}.
One may easily express $\epsilon(\mu)$ by the dimensionless
eigenvalues $a_{nl}$ of the equation
\begin{equation}\label{eig}
- \,\psi(\rho)'' -\frac{1}{\rho}\psi(\rho)'
        + \(\frac{l^2}{\rho^2} + \rho \)\psi(\rho)
\;=\; a_{nl} \psi(\rho)
\end{equation}
through
\begin{equation}\label{eps}
\epsilon_{nl}(\mu) \;=\; \frac{\sigma^{2/3}}{\mu^{1/3}} \,a_{nl} \quad.
\end{equation}
Here $n$ is the principal quantum number and $l$ the orbital one.
The numerical values of $a_{nl}$ for the lowest quantum numbers
have been obtained numerically. They are listed in table 1.
In order to compare with the potential and with the lattice size the
argument of the wave functions has been scaled back in units  $(g^2T)^{-1}$.
$~\psi_{n=1,l=0}~$ and $~\psi_{n=1,l=1}~$, the (radial symmetric) solutions
of the Schr\"odinger equation (\ref{schr}),
are plotted together with the potential in Fig.~\ref{boundstates} versus
$~r=\sqrt{u_1^2+u_2^2}$.

\begin{table}[t]\label{TableEigen}
\begin{center}
\begin{tabular}{|r|lll|}
\hline
$a_{nl}$ & $l$=0 & $l$=1 &  $l$=2 \\
\hline
$n$=1 & 1.74  & 2.87  &  3.82 \\
$n$=2 & 3.67  & 4.49  &  5.26 \\
\hline
\end{tabular}
\end{center}
\caption{The dimensionless eigenvalues $a_{nl}$ of eq.~(\protect\ref{eig})}
\end{table}

\begin{figure}[t]
\begin{picture}(14.0,7.5)
\put(1.0,6.4){\mbox{\LARGE $\frac{M_{nl}}{\sqrt{\sigma}}$}}
\put(7.2,0.0){$m_3^2/\sigma$}
\put(7.4,6.6){\small 2p}
\put(7.4,6.1){\small 1d}
\put(7.4,5.43){\small 2s}
\put(7.4,4.3){\small 1p}
\put(7.4,2.55){\small 1s}
\put(0.5,-0.5){
\epsfxsize13cm \epsffile{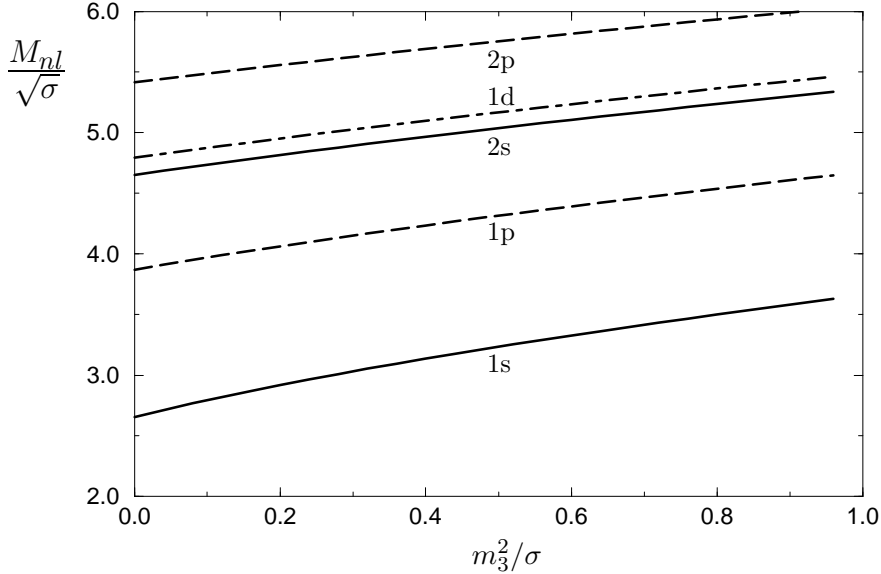}}
\end{picture}
\caption{The masses of the bound states vs.~the squared Lagrangian mass.
The full lines are the masses of the s-states, the dashed lines
are those of the p-states and the dot-dashed line corresponds to the 1d-state}
\label{termschema}
\end{figure}

\begin{figure}[t]
\begin{picture}(14.0,7.5)
\put(6.8,0.0){$r/(g^2T)^{-1}$}
\put(0.7,6.5){\mbox{\LARGE $\frac{V(r)}{g^2T}$}}
\put(4.6,6.05){\footnotesize 1p-wave}
\put(4.6,5.1){\footnotesize 1s-wave}
\put(12.6,5.67){$\epsilon_{11}$}
\put(12.6,4.57){$\epsilon_{10}$}
\put(0.5,-0.5){
\epsfxsize13cm  \epsffile{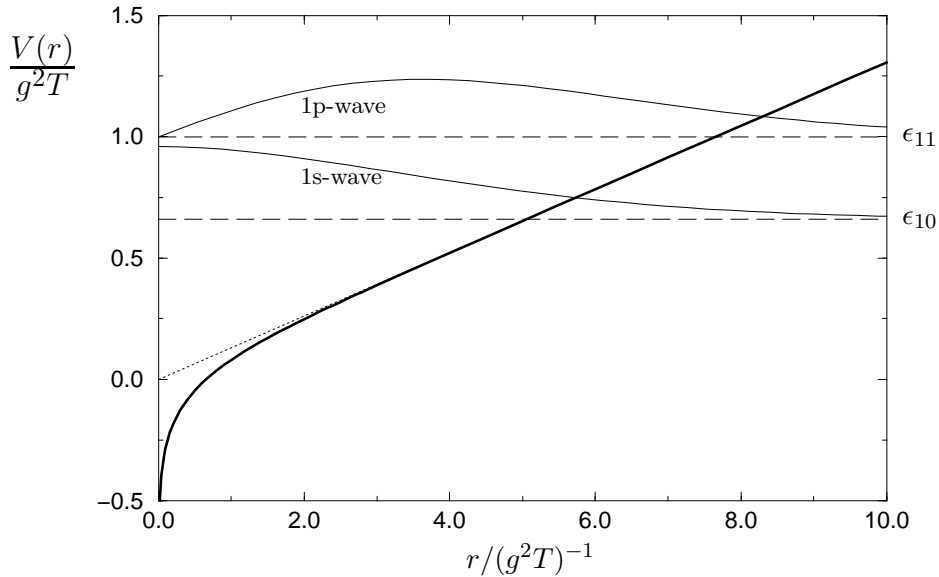}}
\end{picture}
\caption{The potential and the bound states. The thick full line
shows the complete potential while the dotted line is its linear contribution.
The dashed lines are the energies $\epsilon_{nl}$ of \eq{eps}.
They have been used as zeros
for the bound state wave functions, which are in arbitrary units.
We used $~~\sigma=0.13\,(g^2T)^2~~$ and $~~m_3=0.17\,g^2T~~$ for this plot.}
\label{boundstates}
\end{figure}

{}From the steepest descend equation (\ref{steep}) we obtain
\begin{equation}\label{mu}
\mu = \sqrt{\sigma}\; z(a_{nl} , m_3^2/\sigma)^{3/2} \quad,
\end{equation}
where $z(a,y)$ is a solution of the cubic equation
\be
 z^3 - \frac{1}{3} a\, z - y \;=\; 0
\ee\\[-4ex]
yielding
\begin{eqnarray}
z(a,y) \,=\, \frac{2^{1/3}}{3} a \( 27 y + \sqrt{729 y^2 - 4 a^3}\)^{-1/3}
\!\!+ \frac{1}{3 \; 2^{1/3}}\( 27 y + \sqrt{729 y^2 - 4 a^3}\)^{1/3}.
\end{eqnarray}
The mass $M_{nl}$ of the bound state with the quantum numbers $n$ and $l$
is read off \eq{steep} using \eq{eps} and \eq{mu}
\begin{equation}\label{M}
\frac{M_{nl}(m_3/\sqrt{\sigma})}{\sqrt{\sigma}} \;=\;
4 z^{3/2} \;-\; \frac{2 m_3^2}{\sigma z^{3/2}} \quad.
\end{equation}
$M_{nl}$ of the lowest
bound states is plotted in Figure \ref{termschema} versus the
mass of the fundamental scalar (which corresponds to the current quark
mass in QCD).\\
\rule{0cm}{0cm}

Our results show the expected hierarchy of states, i.e.\
the lowest lying state is an isoscalar S-state (the ``Higgs''), followed
by  an isovector P-state (the ``W-boson'').
They should now be compared with existing lattice data .
These data are given for certain values of lattice parameters
$\beta_G$ and $\beta_H$, where $\beta_H$ essentially fixes
the temperature, and the lattice constant is
\mbox{~~$ a = \frac{4}{\beta_G \,g^2 T} $\rule[-0.2cm]{0.0cm}{0.7cm}~}.
Masses in  units of $g^2 T$ are obtained  by multiplying the lattice ones
with $\frac{\beta_G}{4}$.

We consider lattice data of ref.~\cite{IlgenfEA} corresponding
to  a zero temperature Higgs mass of 35 GeV,
and concentrate on the data point $\beta_G=12$, $\beta_H=0.3412$ where
the static force has been measured as well. The mass values are
{}~~$m_H = (0.65\pm 0.1)\, g^2 T$~~and
{}~~$m_W=(1.75\pm 0.08)\, g^2 T$~, measured on a $30^3$ lattice.
The static force is consistent with
a confining potential with
string tension ~~$\sigma=(0.13\pm 0.02)\, (g^2 T)^2$~~
up to a distance of $4 (g^2 T)^{-1} $ where it has been measured.
At some (large) distances screening should set in, which, however,
is presumable not relevant for the lowest bound states.

For a comparison, we have to estimate the  corresponding Lagrangian
mass $m_3$ of the fundamental scalar. From $\beta_H$, it is obtained as
\be\label{lagrmass}
\(\frac{m_3}{g^2 T}\)^2 = \(\frac{\beta_G}{4}\)^2
  \(\frac{2 (1-2 \beta_R- 3 \beta_H)}{\beta_H}\)_{\rm subtr}
\ee\\[-2.5ex]
with \mbox{~~$\beta_R = \frac{\lambda}{g^2} \frac{\beta_H^2}{\beta_G} ~~$
\rule[-0.3cm]{0.0cm}{0.3cm}}, and
the  subscript indicates that the expression has to be  subtracted
at the critical $~\beta_H~$, which is equal to $0.34138$ in the
present case. This leads to
\mbox{~~$m_3=0.17 g^2 T$~~} at ~~$\beta_H = 0.3412$~.
In \eq{lagrmass} we
assume for simplicity that $m_3$ vanishes at the critical $\beta_H$.
Strictly speaking, $m_3$ depends on an arbitrary normalization scale
$\mu_3$; in deriving \eq{lagrmass} we adopt the natural choice
$~\mu_3 \propto g^2 T~$.

A comparison with
lattice data can now be performed based on equations (\ref{M})
(see also Fig.~\ref{termschema}).
Using the string tension ~~$\sigma=(0.13\pm 0.02)\, (g^2 T)^2$~~
and the Lagrangian mass ~$m_3 = 0.17g^2T$~
from the lattice data one sees that our model yields qualitatively
the right results of a substantial splitting between these states.
The predicted mass of the $W$-boson ($m_W = (1.47 \pm 0.11)\, g^2T$)
agrees quite well with the lattice mass,
in view of the approximations made.
Taking into account finite-$a$ effects the lattice value may come
down \cite{IlgenfEA} improving the agreement.
The mass of the composite Higgs ($m_H = (1.06\pm 0.07)\, g^2T$)
calculated by us is a factor 1.6 larger than the lattice value.

One may also take into account the attractive Coulomb force.
The force measured on the lattice is excellently fitted by
\be
F \;=\; 0.123 \,(g^2 T)^2\, K_1(1.01g^2T\; r)\;+\; 0.13 \,(g^2 T)^2
\ee
(cf.\ Fig.~5 of ref.~\cite{IlgenfEA}).
The potential is obtained from $F$ by integration fixing it
to $~~\sigma r~~$ at large distances.
The first term corresponds to the exchange force of a particle
with mass ~$1.01g^2T$~, the constant part is the string tension.
The influence of the Coulomb force is small: it lowers the mass of
the composite Higgs by only 2\%, that of the $W$ by 0.2\%.
This is due to the fact, that the bound state solutions are
much larger than the range of the Coulomb force (cf.\ Fig.~\ref{boundstates}).
The $r$-dependent part of the force can therefore be neglected in
good accuracy justifying  the use of the area law in \eq{arealaw}.

As far as our
model calculations are concerned various approximations are involved.
Perhaps the most important one is neglecting the quantum fluctuations
of the center-of-mass coordinates $R_k$. This approximation should
work better for heavier (recurrent) bound states.
The deviation of the Higgs correlation
mass may be explained by this.

The lattice results are influenced by finite volume as well as
finite lattice spacing effects. Although finite size investigations
did not show significant effects on the masses one should keep
in mind that the predicted size of the bound states is comparable
to the lattice size, which in this case is $10 (g^2 T)^{-1}$ with periodic
boundary conditions (see Fig.~\ref{boundstates}).

Recurrent states in the isoscalar S-channel would be difficult to
identify because they are predicted in the mass range of the $W$-balls of
pure $SU(2)$ gauge theory ($1.5 g^2 T$) \cite{Teper}. More interesting are
the d-states ($l$=2).  The corresponding $W$-ball states have masses of
the order of $2.5 g^2 T$ which is significantly heavier
than the predicted $l$=2 scalar--anti-scalar bound state mass
$m_{1d}$ of about  $1.8 \, g^2 T$~.
Note, nevertheless, that this is above the threshold of the two Higgs
channel corresponding to the measured masses.
Lattice measurements of this channel are under way.

The predicted masses are based on the evaluation of the Green function
\eq{FSR}, which could only be performed with some approximations.
Even without these approximation one finds that the masses of
the bound state model depend only on the string tension $\sigma$
and the Lagrangian mass $m_3$.
Considering lattice data in the whole $\beta_H$ (i.e.\ $m_3$) range
one finds a stronger $m_3$-dependence of $~m_W/(g^2 T)~$ and  $~m_H/(g^2 T)~$
than we would predict if $~\sqrt{\sigma}/(g^2 T)~$ was constant.
This might indicate that  $~\sqrt{\sigma}/(g^2 T)~$ itself
depends significantly
on $m_3$, dropping as $m_3$ approaches zero.
It would be interesting  to verify or disprove this by lattice studies.
$~\sqrt{\sigma}/(g^2 T)~$ could also show a sizable $\lambda$-dependence
(decreasing with increasing $\lambda$).

One should also keep in mind that the static potential will not continue
to rise linearly at larger distances but charge screening will set in.
The screening length has not been measured so far. As a consequence of
charge screening, heavier bound states will disappear. The mass of a
bound state close to the threshold would  be affected as well.\\
\rule{0cm}{0cm}

In conclusion the hot electroweak phase appears as an intriguing case
of a confining 3-dimensional gauge system. It is similar to
3-dimensional QCD but with a fundamental scalar $\Phi$ instead of quarks
and hence without spin-interactions and without the problems of
chiral symmetry breaking.
The masses of the low lying bound states have been calculated from
the Green function within some approximations. The comparison with
lattice results is satisfactory.
Our discussion is open for refinement  both on the lattice side because of the
limited lattice size and on the side of calculating the masses of
relativistic bound states, a notorious problem also in QCD.
For the latter question also other methods can be envisaged,
including sum rules for the  3-dimensional
Abbott-Farhi-model \cite{SchDoKr}-\cite{BuSch}.
An enlarged version of this letter is in progress \cite{Wir3}.


\end{document}